\documentclass[jkps,preprint,fleqn,showpacs,showkeys]{revtex4}
\usepackage{graphicx}
\usepackage{amssymb}
\usepackage{amsmath}
\usepackage{bm}
\begin{document}
\setcounter{page}{0}
\title[]{Dirac Coupled Channel Analyses of Proton Inelastic Scattering from $^{40}$Ar Nucleus}	
\author{ Sugie \surname{Shim}}
\email{shim@kongju.ac.kr}
\thanks{Fax: +82-41-850-8489}
\affiliation{Department of Physics, Kongju National University, Gongju 314-701}

\date[]{Received 2015}

\begin{abstract}
 0.8 GeV proton inelastic  scatterings from $^{40}$Ar nucleus
 are analyzed using  an optical   potential model in  the Dirac coupled   channel
formalism.
A   rotational  collective model is used to obtain the transition  optical
potentials  for the low  lying excited states of the rotational  band.
The optical potential parameters of a Woods-Saxon shape and the deformation parameters of the excited states are searched phenomenologically  to reproduce the experimental differential cross-section data by using the sequential iteration method.
The effect of a multistep  process is investigated by including the channel
coupling between two excited  states in addition to the couplings between the ground
state and the excited states.
 The calculated deformation parameters of the excited states are compared with those obtained by using the nonrelativistic calculations.
 The results of the Dirac coupled channel calculations are found to show pretty good agreements with the  experimental data of the ground state and
the low-lying excited states. The multistep excitation process via the channel coupling with the second $2^+$ state is found to be important for the excitation of the $4^+$ state at the inelastic scatterings from the deformed nucleus, $^{40}$Ar.

\end{abstract}

\pacs{25.40.Ep, 24.10.Jv, 24.10.Ht, 24.10.Eq, 21.60.Ev}

\keywords{Dirac phenomenology, Coupled channel calculation, Optical potential model, Collective model, Proton inelastic scattering}

\maketitle

\section{INTRODUCTION}

Relativistic approaches based on the Dirac equation as the relevant wave equation have proved to be very successful in treating
the proton elastic and inelastic scatterings at intermediate
energies  from the spherically  symmetric nuclei [1-4] and a few deformed  nuclei [5-12].  Considerable improvements have been shown in the
coupled channel calculations using the Dirac phenomenology compared to  the  classical nonrelativistic calculations based on the Schr\"{o}dinger equation \cite{8,9,10,11,12,13}.

   In this work we use phenomenological optical potentials in the Dirac coupled  channel
calculation [1,2] to analyze 0.8 GeV proton  inelastic scatterings  from a  deformed
  nucleus, $^{40}$Ar. So far only  the distorted wave  Born  approximation (DWBA)
calculation using  phenomenological optical  potentials  in  nonrelativistic
formalism neglecting  the coupled  channel effects, or the nonrelativistic coupled channel calculation at low energy  has been done  to analyze the
inelastic scatterings of proton from $^{40}$Ar nucleus [14, 15].
A rotational collective model  is used for the  transition optical
potentials to accommodate the collective  motion of excited deformed nuclei
considering the low lying excited states of  the rotational bands [5].
The multistep excitation process is included  in the calculation by considering
the  coupling between two excited states, in  addition to the couplings
between the
ground and the excited states.
In order to solve the complicated Dirac coupled channel equations, we use a computer program called ECIS \cite{16} where the Dirac optical potential and the deformation parameters are determined phenomenologically by using a sequential iteration method. The Dirac equations are reduced to Schr\"{o}dinger-like second-order differential equations to obtain the effective central and spin-orbit optical potentials, and the obtained effective potentials are analyzed and compared with those of the nonrelativistic calculations. The calculated deformation parameters for the low-lying excited states of the $^{40}$Ar nucleus are analyzed and compared with those calculated by using the nonrelativistic approaches.

\section{Theory and Results}

Dirac analyses are performed phenomenologically for 0.8 GeV unpolarized proton inelastic scatterings from $^{40}$Ar nucleus by using an optical potential model and a collective model.
Because $^{40}$Ar is a spin-0 nucleus,  only scalar,  time-like  vector  and  tensor
optical potentials  survive \cite{17, 18}, as in spherically-symmetric nuclei; hence, the relevant Dirac equation for the elastic scattering from the nucleus is given as
\begin{equation}
[\alpha \cdot p + \beta ( m + U_S ) - ( E - U_V^0- V_c )
+ i \alpha \cdot  \hat{r} \beta U_T ] \Psi(r) = 0 .
\label{e1}
\end{equation}
Here, $U_S$ is a scalar potential, $U_V^0$ is a time-like vector potential, $U_T$ is a tensor potential, and $V_c $ is the Coulomb potential.
Even  though tensor  potentials  always  present due  to the
interaction  of the  anomalous magnetic  moment  of the  projectile with the
charge distribution of the  target, they have been found to  be always very
small compared  to the scalar or the vector potentials \cite{2}.  Hence they  are neglected in
this calculation.
 The scalar and the vector optical potentials are complex, and have the Woods-Saxon shapes \cite{13} as they are assumed to follow the distribution of the nuclear density.
In the first-order rotational model of ECIS, the deformation of the radius of the optical potential is given by using the Legendre polynomial expansion method \cite{13}.
We assume that the shapes of the deformed potentials follow the shapes of the deformed nuclear densities and that the transition potentials can be obtained by assuming that they are proportional to the first-order derivatives of the diagonal potentials. However, depending on the  model assumed, pseudo-scalar and
axial-vector potentials may also be present in the equation when we consider the inelastic
scattering. In the collective model approach used in this work, we assume that we can obtain appropriate transition potentials by deforming the  direct potentials that describe the elastic
channel reasonably well \cite{19}.
The transition potentials are given by
\begin{equation}
U_S ^\lambda (r) = {\beta_S^\lambda \over (2\lambda+1)^{1/2}} \; \left [
R_S^r {d(ReU_S (r)) \over dR_S^r} + i R_S^i  {d(ImU_S (r)) \over dR_S^i}
\right ]
 \; Y_{\lambda 0}^*  (\Omega),
\end{equation}
\begin{equation}
U_V ^{0,\lambda} (r) =
{\beta_V^\lambda \over (2\lambda+1)^{1/2}} \; \left [ R_V^r
 {d(ReU_V^0 (r)) \over dR_V^r} +  i R_V^i {d(ImU_V^0 (r)) \over dR_V^i}  \right ] \;
Y_{\lambda 0}^*  (\Omega),
\end{equation}
 where $\lambda$ is the multipolarity, superscripts $r$ and $i$ refer to the real
and the imaginary parts of the radius ($R$) parameters.  The real and the imaginary $\beta_\lambda$ are
taken to be  equal for a given  potential type,  so that $\beta_S$ and
$\beta_V$ are
determined for  each excited state. The Dirac coupled  channel equations
are  solved
numerically using  the computer code  ECIS [16]  which employs the sequential
iteration method.
We consider  both the  couplings  between the  $0^+$ ground  state  and the low
lying excited   states and   also coupling    between two   excited states,
for  example, between the $2^+$ and the $4^+$ states. Hence the multistep process is
included in  the calculation.  Because  the channel  couplings  between the
low-lying excited
states of the rotational band are strong  in the inelastic scatterings from  a deformed
nucleus,
the  multistep transition process could be important.
In order to compare the calculated results with those of nonrelativistic calculations, we reduce the Dirac equation to a Schr\"{o}dinger-like second-order differential equation by considering the upper component of the Dirac wave function and obtain the effective central and spin-orbit optical potentials \cite{2}.

 The  elastic and
inelastic  data are obtained from Ref. 14 for 0.8 GeV proton inelastic scatterings from $^{40}$Ar nucleus. The  low lying excited states of the $2^+$ (1.461 MeV), $2^+$ (2.524 MeV) and $4^+$ (2.89 MeV) are
considered and assumed  to be collective  rotational states in  the calculation. The 12 parameters of the
  diagonal scalar and  vector optical  potentials in the Woods-Saxon shapes are obtained by fitting the elastic scattering data.
  Pretty good agreements  with the elastic scattering  data are obtained  as  shown  by the dashed  lines in Fig. 1. For  the
inelastic  scattering calculations, we include  only the ground and one excited
states  at  once in the calculations as a  first step. Next,  the
ground, the first $2^+$ or the second $2^+$, and $4^+$ states are  included in the calculations  to investigate
the   effect of the channel coupling between the excited states.

\begin{figure}
\includegraphics[width=10.0cm]{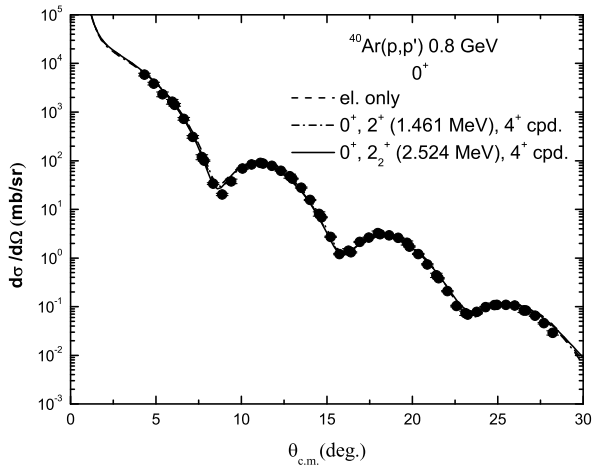}
\caption{Differential cross section of the ground state for 0.8 GeV p +  $^{40}$Ar scattering. The dashed, the dash-dot and the solid lines represent the results of Dirac phenomenological calculations where only the ground state is considered, where the ground, the first $2^+$ (1.461 MeV) and the $4^+$ states are coupled, and where the ground, the second $2^+$ (2.524 MeV) and the $4^+$ states are coupled, respectively.}
\label{fig1}
\end{figure}

\begin{figure}
\includegraphics[width=10.0cm]{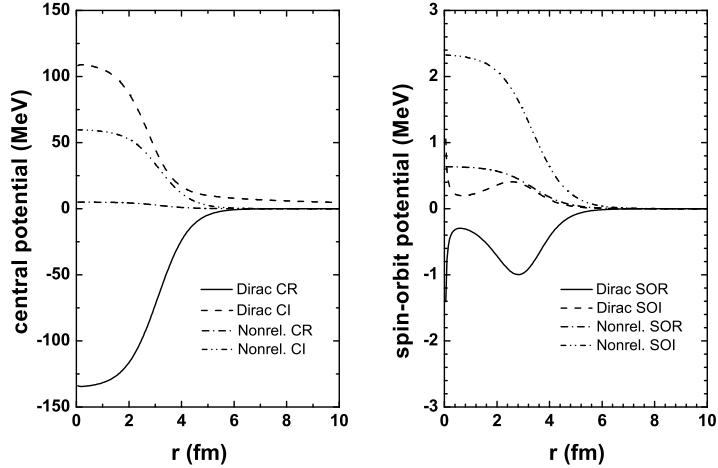}
\caption{Comparison of the Dirac effective central and spin-orbit potentials of $^{40}$Ar with those of the nonrelativistic (Nonrel.) calculations. CR, CI, SOR and SOI mean central real, central imaginary, spin-orbit real, and spin-orbit imaginary potentials, respectively.}
\label{fig2}

\end{figure}

\begin{table}
\caption{Calculated phenomenological optical potential parameters of a Woods-Saxon shape for 0.8 GeV proton elastic scatterings from $^{40}$Ar nucleus.}
\begin{ruledtabular}
\begin{tabular}{cccccccc}

   Potential      &   Strength (MeV)   & Radius (fm)  & Diffusiveness (fm)  ~ \\
   \hline
 Scalar  & -472.8     & 3.154 & 0.5967        ~ \\
 real    &      &     &            ~ \\ \hline
 Scalar  & 120.3     & 3.013 & 0.6896        ~ \\
 imaginary    &      &     &            ~ \\ \hline
 Vector  & 264.0    & 3.153 & 0.5954         ~ \\
 real    &      &     &           ~ \\ \hline
 Vector  & -127.3     & 3.129 & 0.6254       ~ \\
 imaginary  &      &     &             ~ \\
\end{tabular}
\end{ruledtabular}
\label{table1}
\end{table}

The calculated optical potential parameters of the Woods-Saxon shape for 0.8 GeV proton elastic scatterings from $^{40}$Ar nucleus are shown in Table 1. $\chi^2/N$, where $N$ is the number of experimental data, is about 13.4.
We observe that the real parts of the scalar potentials and the imaginary parts of the vector potentials turn out to be large and negative and that the imaginary parts of the scalar potentials and the real parts of the vector potentials turn out to be large and positive, showing the same pattern as in spherically-symmetric nuclei \cite{2, 3, 4}.

In Fig. 2, the Dirac effective central and spin-orbit potentials for $^{40}$Ar are compared with those of the nonrelativistic calculations \cite{14}. We should note that one of the merits of using the relativistic approach based on the Dirac equation instead of using the nonrelativistic approach based on the Schr\"{o}dinger equation is that the spin-orbit potential appears naturally in the Dirac approach when the Dirac equation is reduced to a Schr\"{o}dinger-like second-order differential equation, whereas the spin-orbit potential must be inserted by hand in the nonrelativistic Schr\"{o}dinger approach.
The surface-peaked phenomena observed for most of the s-d shell deformed nuclei \cite{7,10,11,12,13} are not observed at the effective central potentials when $^{40}$Ar nucleus is considered.
 The strength of the real effective central potential of the Dirac approach turns out to be large, about -130 MeV at the center of the nucleus, compared to that of the nonrelativistic Schr\"{o}dinger approach, which is about 4.0MeV \cite{14}.
 The real parts of the effective central and spin-orbit potentials turn out to have negative values as in the cases of the other deformed nuclei \cite{12, 13}, while those of the nonrelativistic calculations are found to have positive values.
Surface-peaked phenomena are clearly shown at the effective spin-orbit potentials, indicating that the spin-orbit interaction is a surface-peaked interaction.
The surface-peaked shape cannot be obtained in the conventional nonrelativistic calculations because the Woods-Saxon shape is used for both the central and the spin-orbit potentials, as shown in the figure.

 \begin{figure}
\includegraphics[width=10.0cm]{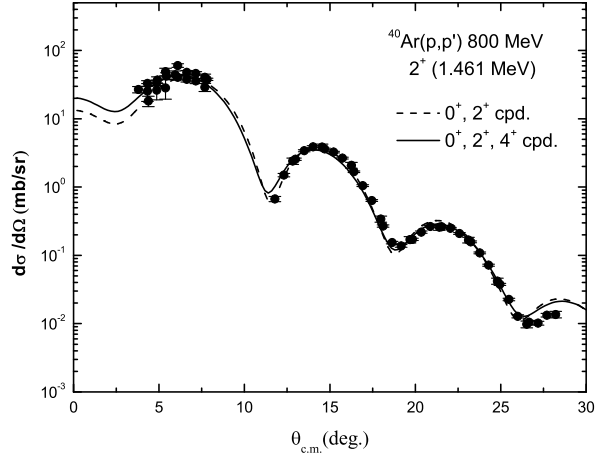}
\caption{Differential cross section of the first $2^+ $ state (1.461 MeV) for 0.8 GeV p +  $^{40}$Ar inelastic scattering. The dashed and the solid lines represent the results of Dirac coupled channel calculations where the ground and the first $2^+$ states are coupled and where the ground, the first $2^+$ and the $4^+$ states are coupled, respectively.}
\label{fig3}
\end{figure}

\begin{figure}
\includegraphics[width=10.0cm]{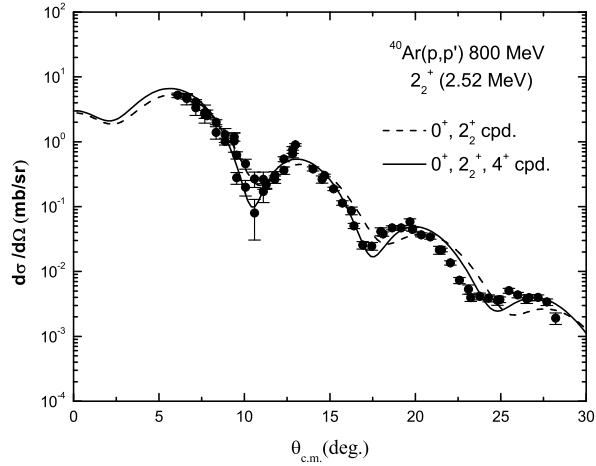}
\caption{Differential  cross section of the second $2^+ $ state (2.524 MeV) for 0.8 GeV p +  $^{40}$Ar inelastic scattering. The dashed and the solid lines represent the results of Dirac coupled channel calculations where the ground and the second $2^+$ states are coupled and where the ground, the second $2^+$ and the $4^+$ states are coupled, respectively.}
\label{fig4}
\end{figure}

\begin{figure}
\includegraphics[width=10.0cm]{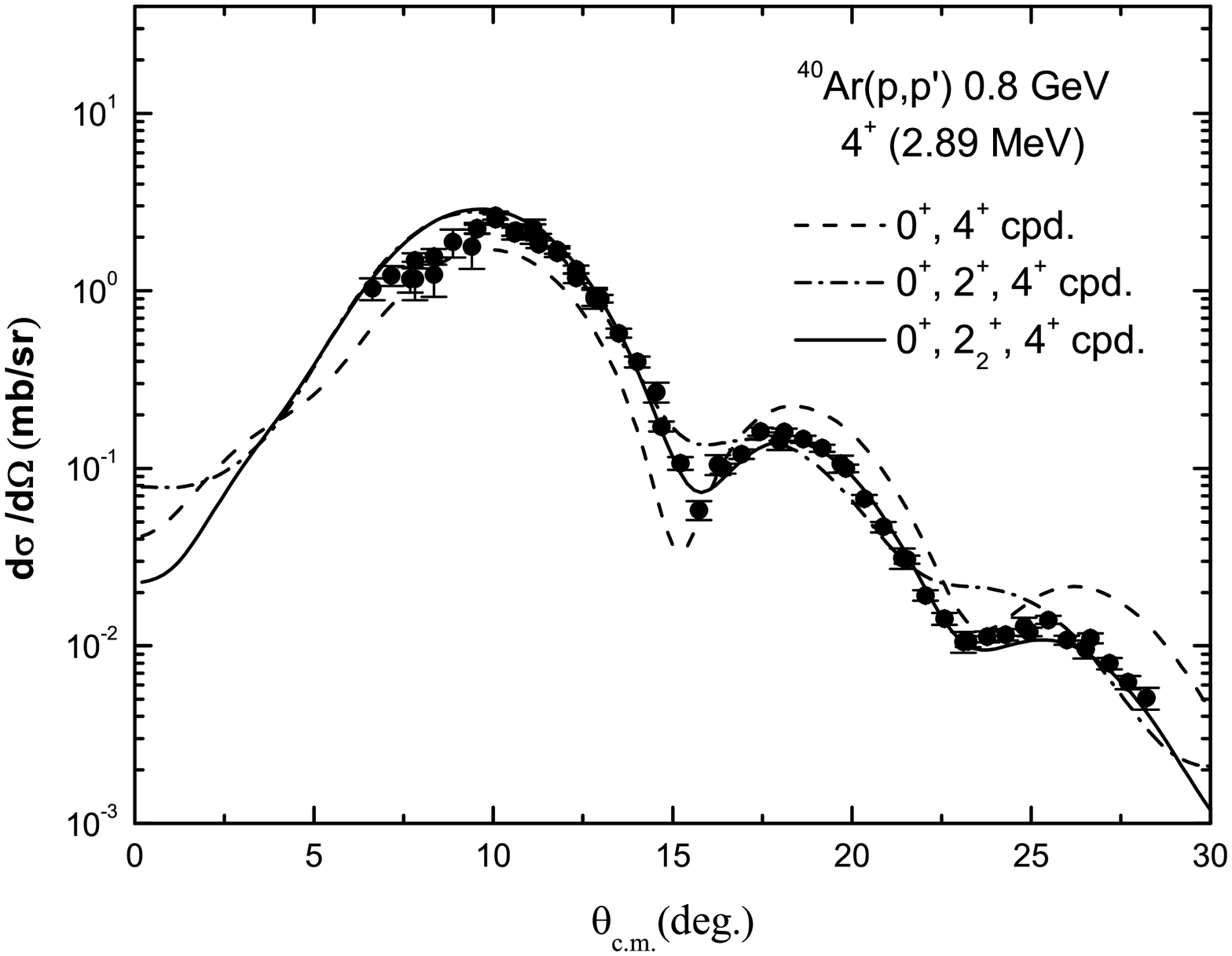}
\caption{Differential  cross section of the $4^+ $ state for 0.8 GeV p +  $^{40}$Ar inelastic scattering. The dashed, the dash-dot and the solid lines represent the results of Dirac coupled channel calculations where the ground and the $4^+$ states are coupled, where the ground, the first $2^+$ and the $4^+$ states are coupled, and where the ground, the second $2^+$ and the $4^+$ states are coupled, respectively.}
\label{fig5}
\end{figure}

Next, six-parameter searches are performed by including one excited state in addition to the ground state, starting from the 12 parameters obtained for the direct optical potentials in the elastic scattering calculation. Here, six parameters are the two deformation parameters, $\beta_S$ and $\beta_V$, for the each excited state and the four potential strengths, the scalar real and imaginary potential strengths and the vector real and imaginary potential strengths, keeping the potential geometries unchanged. The optical potential strengths obtained by fitting the elastic scattering data in the elastic scattering calculation are varied because the channel coupling of the excited states to the ground state should be included in the inelastic scattering calculation.
Finally, eight-parameter searches are performed by considering three states, the ground, the first $2^+$ or the second $2^+$ state, and the $4^+$ states, together in the calculation in order to investigate the effect of the channel coupling between the excited states, and the results are compared with those obtained by the calculation where only the ground and one excited states are coupled.
Figure 1 shows the results of the coupled channel calculations for the ground state, showing about the same pretty good agreements with the experimental elastic differential cross section data in all the cases. In the figures, `cpd' means `coupled'. In Figs. 3, 4 and 5, the calculated observables for the first $2^+$, the second $2^+$, and the $4^+$ states are compared with the experimental data. For the first $2^+$ state, the agreements with the experimental data are pretty good even when only the ground and the first $2^+$ states are coupled as shown by the dashed lines in Fig. 3. The agreements with the experimental data are not changed much by adding the coupling with  the $4^+$ state, as shown by the solid lines. $\chi^2/N$ for the two cases turns out to be about the same. The agreements with the experimental data for the $4^+$ state are not good when only the ground and the $4^+$ states are coupled, as shown by the dashed lines in Fig. 5 and they are not improved much by adding the coupling with the first $2^+$ state, as shown by the dash-dot lines.
However, the agreements with the experimental data for the $4^+$ state are improved significantly by adding the coupling with the second $2^+$ state, as shown by the solid lines in Fig. 5, indicating that the multistep excitation process via the channel coupling with the second $2^+$ state is important for the excitation of the $4^+$ state. The agreements with the experimental data for the second $2^+$ state are also improved noticeably by adding the coupling with the $4^+$ state, as shown by the solid lines in Fig. 4. Hence the multistep transition process is confirmed  to be important because the excited states of a rotational band, they are the second $2^+$ and the $4^+$ states in this case,
are strongly coupled to each other, as shown previously for the inelastic scatterings from the other axially-symmetric deformed nuclei \cite{10, 11, 12, 13}.
This is not the case for the spherically-symmetric nuclei, where the excited states
 are well described by considering the coupling via
single-step  transitions \cite{2, 3}. Even though the results of the Dirac coupled channel calculation show pretty good agreements with the second $2^+$ data, the second and the third minima of the diffraction pattern are found to be shifted slightly from the data.
This slight discrepancy could be due to the coupling effect of the second $2^+$ state with the excited $0^+$ state at 2.12 MeV, that is omitted in the calculation.
Reproducing hadron scattering experimental data of excited $0^+$ levels is known to be notoriously difficult because the form factors of $\lambda=0$ transitions, the assignment of the excited $0^+$ state to a band structure,
the reduced matrix elements, and the relative phases for all the multipolarities of the different transitions are not well known \cite{15}. Hence the coupling to the excited $0^+$ state is not included in the calculation,
even though the excited $0^+$ state, the second $2^+$ state and the $4^+$ state could form a $\beta$ rotational band \cite{14, 15} and coupled to each other.
The potential strengths are changed to -185.1, 396.0, 117.8, and -264.6 MeV for the scalar real and imaginary and the vector real and imaginary potentials, respectively, in the first $2^+$ state coupled case, to -479.1, 109.2, 273.3, and -122.6 MeV in the second $2^+$ state coupled case, to -383.9, 302.8, 219.5, and -206.7 MeV in the $4^+$ state coupled case, to -462.9, 191.9, 264.1, and -157.5 MeV in the first $2^+$ and the $4^+$ states coupled case, and to -502.3, 93.41, 283.2, and -114.4 MeV in the second $2^+$ and the $4^+$ states coupled case.
The results of our relativistic coupled channel calculation show clearly better agreements with the experimental data than those of the nonrelativistic DWBA calculations \cite{14}.

\begin{table}
\caption{Comparison of the deformation parameters for the first $2^+ $, the second $2^+ $ and the $4^+$ excited states for 0.8 GeV proton inelastic scatterings from $^{40}$Ar nucleus with those obtained by using the nonrelativistic calculations.}
\begin{ruledtabular}
\begin{tabular}{c|ccccccc}
   & Energy   &   &   &   ~ \\
   &   (MeV)  & $\beta_S $  & $\beta_V $  & $\beta_{NR} $  ~ \\
   \hline \hline
 $2^+ $ state  &  1.461  &   .28   &  .27   &  $.28^{14} $  ~ \\
 $2^+_2 $ state  &  2.524  &   .085   &  .058   &  $.082^{14} $  ~ \\
 $4^+ $ state  &  2.89  &   .092   &  .057   &  $.0935^{14} $       ~ \\

\end{tabular}
\end{ruledtabular}
\label{table2}
\end{table}

In Table 2, we show the deformation parameters for the first $2^+$, the second $2^+$ and the $4^+$ excited states of $^{40}$Ar nucleus. $\beta_S$ is observed to be always larger than $\beta_V$ for the excited states that we considered in the calculations.  The deformation parameters obtained from the Dirac phenomenological coupled channel calculation for the first $2^+$, the second $2^+$ and the $4^+$ excited states of $^{40}$Ar are found to show pretty good agreements with those obtained by using the nonrelativistic DWBA calculations \cite{14} using the same Woods-Saxon shapes for the geometries of the optical potentials, even though the theoretical bases are quite different.

\section{CONCLUSIONS}

A relativistic Dirac coupled channel calculation using an optical potential model is able to  describe the low-lying excited states of the rotational bands for 0.8 GeV unpolarized proton inelastic scatterings from a deformed nucleus $^{40}$Ar pretty well. The observables obtained by our relativistic coupled channel calculation show clearly better agreements with the experimental data than those obtained by the nonrelativistic DWBA calculations.
 The Dirac equations are reduced to second-order differential equations to obtain Schr\"{o}dinger-equivalent effective central and spin-orbit potentials, and surface-peaked phenomena are observed only at the effective spin-orbit potentials for the scatterings from $^{40}$Ar nucleus. The first-order rotational collective models are used to describe the low-lying excited states of the rotational band in the nucleus, and the calculated deformation parameters are compared with those obtained by using the nonrelativistic calculations.
  The deformation parameters obtained from the Dirac phenomenological calculations for the first $2^+$, the second $2^+$, and the $4^+$ state of $^{40}$Ar are found to show pretty good agreements with those obtained by using the nonrelativistic calculations.
  The agreements with the experimental data for the $4^+$ state are not good when only the ground and the $4^+$ states are coupled or when the ground, the first $2^+$ and the $4^+$ states are coupled. But they are improved significantly when the ground, the second $2^+$ and the $4^+$ states are coupled, indicating that the multistep excitation process via the channel coupling with the second $2^+$ state is important for the excitation of the $4^+$ state for the inelastic scattering from the deformed nucleus, $^{40}$Ar.

\end{document}